\documentclass[aps,prl,twocolumn,reprint,superscriptaddress]{revtex4-1}  % for double-spaced reprint
\usepackage{graphicx}  % needed for figures
\usepackage{dcolumn}   % needed for some tables
\usepackage{bm}        % for math
\usepackage{amssymb}   % for math
\usepackage{amsmath}

\usepackage[mathlines]{lineno}% Enable numbering of text and display math
\begin{document}

% title
\title{Measurement of transverse wakefields induced by a \\ misaligned positron bunch in a hollow channel plasma accelerator}

% authors
\author{C. A. Lindstr\o m}
\email{c.a.lindstrom@fys.uio.no}
\affiliation{Department of Physics, University of Oslo, 0316 Oslo, Norway}
\author{E. Adli}
\affiliation{Department of Physics, University of Oslo, 0316 Oslo, Norway}
\author{J. M. Allen}
\affiliation{SLAC National Accelerator Laboratory, Menlo Park, California 94025, USA}
\author{W. An}
\affiliation{Department of Electrical Engineering, University of California-Los Angeles, Los Angeles, California 90095, USA}
\author{C. Beekman}
\affiliation{LOA, ENSTA ParisTech, CNRS, Ecole Polytechnique, Universit\'{e} Paris-Saclay, 91762 Palaiseau, France}
\author{C. I. Clarke}
\affiliation{SLAC National Accelerator Laboratory, Menlo Park, California 94025, USA}
\author{C. E. Clayton}
\affiliation{Department of Electrical Engineering, University of California-Los Angeles, Los Angeles, California 90095, USA}
\author{S. Corde}
\affiliation{LOA, ENSTA ParisTech, CNRS, Ecole Polytechnique, Universit\'{e} Paris-Saclay, 91762 Palaiseau, France}
\author{A. Doche}
\affiliation{LOA, ENSTA ParisTech, CNRS, Ecole Polytechnique, Universit\'{e} Paris-Saclay, 91762 Palaiseau, France}
\author{J. Frederico}
\affiliation{SLAC National Accelerator Laboratory, Menlo Park, California 94025, USA}
\author{S. J. Gessner}
\altaffiliation[Now at ]{CERN, Geneva, Switzerland.}
\affiliation{SLAC National Accelerator Laboratory, Menlo Park, California 94025, USA}
\author{S. Z. Green}
\affiliation{SLAC National Accelerator Laboratory, Menlo Park, California 94025, USA}
\author{M. J. Hogan}
\affiliation{SLAC National Accelerator Laboratory, Menlo Park, California 94025, USA}
\author{C. Joshi}
\affiliation{Department of Electrical Engineering, University of California-Los Angeles, Los Angeles, California 90095, USA}
\author{M. Litos}
\affiliation{Department of Physics, University of Colorado Boulder, Boulder, CO 80309, USA}
\author{W. Lu}
\affiliation{IFSA Collaborative Innovation Center, Department of Engineering Physics, Tsinghua University, Beijing 100084, China}
\author{K. A. Marsh}
\affiliation{Department of Electrical Engineering, University of California-Los Angeles, Los Angeles, California 90095, USA}
\author{W. B. Mori}
\affiliation{Department of Physics and Astronomy, University of California-Los Angeles, Los Angeles, California 90095, USA}
\author{B. D. O'Shea}
\affiliation{SLAC National Accelerator Laboratory, Menlo Park, California 94025, USA}
\author{N. Vafaei-Najafabadi}
\altaffiliation[Now at ]{Department of Physics and Astronomy, Stony Brook University, Stony Brook, New York 11794, USA.}
\affiliation{Department of Electrical Engineering, University of California-Los Angeles, Los Angeles, California 90095, USA}
\author{V. Yakimenko}
\affiliation{SLAC National Accelerator Laboratory, Menlo Park, California 94025, USA}

% date
%\date{\today}

% abstract
\begin{abstract}
Hollow channel plasma wakefield acceleration is a proposed method to provide high acceleration gradients for electrons and positrons alike: a key to future lepton colliders. However, beams which are misaligned from the channel axis induce strong transverse wakefields, deflecting beams and reducing the collider luminosity. This undesirable consequence sets a tight constraint on the alignment accuracy of the beam propagating through the channel. Direct measurements of beam misalignment-induced transverse wakefields are therefore essential for designing mitigation strategies. We present the first quantitative measurements of transverse wakefields in a hollow plasma channel, induced by an off-axis 20 GeV positron bunch, and measured with another 20 GeV lower charge trailing positron probe bunch. The measurements are largely consistent with theory.
\end{abstract}

\maketitle

%--- INTRODUCTION ---%
Precision tests of the Standard Model of particle physics can be performed with a linear electron-positron collider. However these machines will be very large and expensive to build. Plasma wakefield acceleration (PWFA) \cite{ChenPRL1985,RuthPA1985,JoshiPT2003} is a promising new technique for building a more compact, more cost-effective accelerator: an intense charged particle bunch is propagated through a uniform plasma, where it induces a highly nonlinear wake structure with strong accelerating and focusing fields. While this mechanism has been shown to sustain large acceleration gradients \cite{BlumenfeldNature2007} and high energy transfer efficiency \cite{LitosNature2014} for a second trailing electron bunch, the success does not immediately extend to positrons due to the inherently charge-asymmetric response of nonlinear plasmas. Positron bunches have been transported through and accelerated by meter long plasma wakes \cite{HoganPRL2003,BluePRL2003,CordeNature2015,DocheSciRep2017}. However the extremely nonlinear focusing fields of such wakes make it very difficult to preserve the emittance of the accelerating beam \cite{MuggliPRL2008}.

A possible solution for symmetrizing the acceleration of electrons and positrons while preserving the emittance is to use a hollow channel surrounded by an annular plasma \cite{ChiouPP1995,SchroederPRL1999,KimuraPRAB2011}. This is so because a drive bunch propagating exactly on the channel axis drives an oscillating longitudinal wakefield that moves synchronously with the beam and is transversely uniform, while the transverse (deflecting) wakefield is zero everywhere in the channel. This method \cite{GessnerNComms2016} has been experimentally demonstrated to accelerate positrons \cite{GessnerThesis2016}. However, if the bunch propagates off-axis, it is expected to induce a strong dipole-like transverse wakefield that deflects both the drive beam and the accelerating trailing beam away from the axis. This leads to significantly reduced collider luminosity or even beam loss.

%--- FIGURE 1: Experimental setup ---%
\begin{figure*}[t]
	\centering\includegraphics[width=0.95\textwidth]{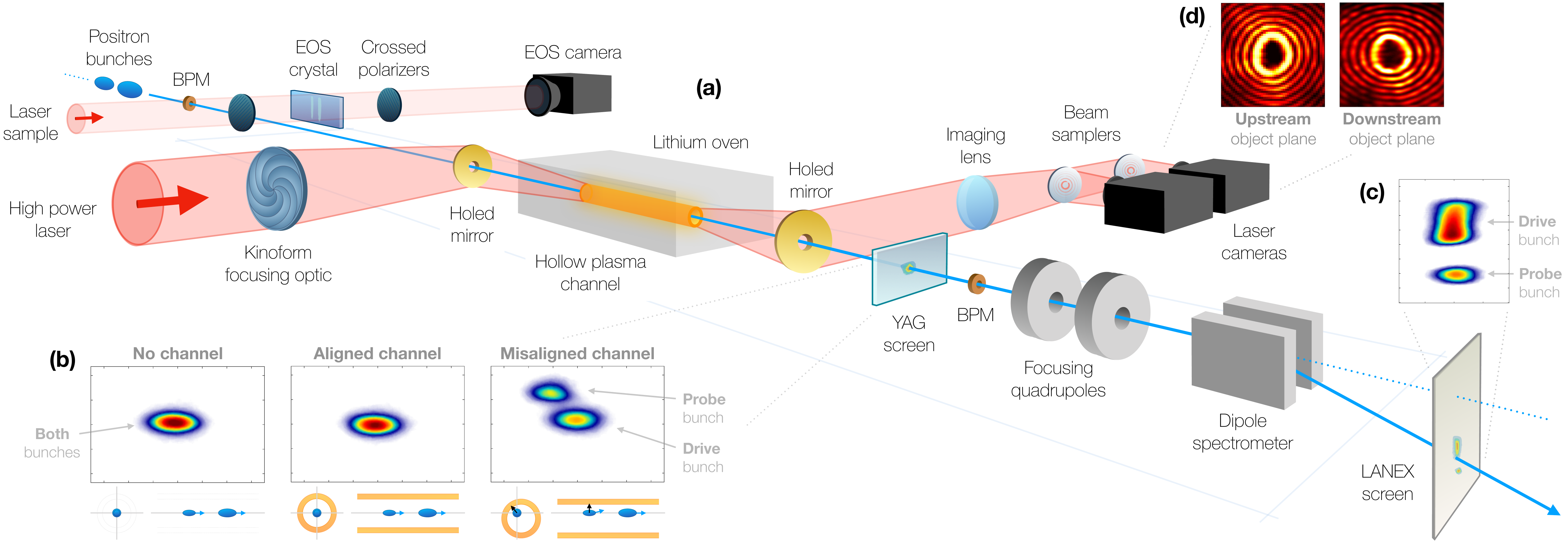}
	\caption{(a) Experimental setup: Two positron bunches first pass an electro-optical sampler (EOS). A Ti:Sapphire laser focused with a kinoform into a lithium vapor oven produces the hollow plasma channel. Two beam position monitors (BPM) measure the trajectory of the beam and an yttrium aluminum garnet (YAG) screen is used to measure the transverse profile (b). A dipole spectrometer with two quadrupoles focuses the beam onto a LANEX screen for energy and angular deflection measurements (c) before it is dumped. Meanwhile, the outgoing laser pulse is focused onto cameras imaging the kinoform profile (d) at different object planes inside the channel, which appears asymmetric due to aberrations induced by the transmissive optics. The upstream spectrometer does not appear in this figure.}
    \label{fig:ExptSetup}
\end{figure*}

%--- EXPERIMENTAL SETUP ---%
In this letter, we present the first experimental measurements of transverse wakefields in a hollow channel plasma accelerator, performed at the Facility for Advanced aCcelerator Experimental Tests (FACET) \cite{HoganNJP2010} at SLAC National Accelerator Laboratory. The plasma channel was formed by ionizing lithium vapor with the high power FACET laser \cite{GreenPPCF2014}, which delivered a maximum of 10~mJ on target in as little as 50~fs (full width at half maximum). A high-order Bessel intensity profile (${J_7}^2$) with the first maximum at 250~$\mu$m was obtained using a kinoform optic that focused the laser close to the center of a 46~cm heat-pipe oven \cite{MuggliIEEE1999}, giving a $25\pm1$~cm long hollow channel. The vapor pressure was set to 3.4~Torr at temperature 1095~K, giving a neutral vapor density of $3\times10^{16}$~cm$^{-3}$. The laser pulse energy was attenuated to ionize only the channel wall, ensuring a truly zero plasma density on axis. A 20.35~GeV two-bunch positron beam was synchronized to arrive a few picoseconds after the laser pulse. The two bunches were obtained from a single bunch by giving it a head-to-tail energy chirp and energetically dispersing it onto a beam notching device, allowing a tuneable bunch separation up to 600~$\mu$m. The positron beam was focused at the channel center with rms beam sizes $\sigma_x=35$~$\mu$m and $\sigma_y=25$~$\mu$m and beta functions $\beta_x=0.5$~m and $\beta_y=5$~m, which ensured that the beam size was approximately constant throughout the channel. A total charge of $0.51\pm0.04$~nC, sufficiently low to not ionize the on-axis lithium vapor, was distributed between the leading drive bunch and the trailing probe bunch with a ratio $(4.1\pm1.1):1$.

%--- EXPERIMENT and INSTRUMENTATION ---%
The experiment consisted of measuring the transverse wakefield in a hollow plasma channel by observing the angular deflection of the probe bunch caused by an offset channel. In particular, the longitudinal variation of the transverse wakefield was measured by means of a bunch separation scan. Figure \ref{fig:ExptSetup} shows the experimental setup. Although the two bunches originated from one bunch, scanning the bunch separation was possible by stretching the bunch and adjusting the beam notching device \cite{LitosNature2014}. An electro-optical sampler (EOS) was used to measure the longitudinal bunch profile of the incoming beam, and an yttrium aluminum garnet (YAG) crystal in a horizontally dispersive region functioned as an upstream energy spectrometer for the positron beam. Two beam position monitors (BPMs) were used to measure the beam trajectory. Downstream of the channel, a non-destructive YAG screen was used to measure the transverse profile of the outgoing beam. A spectrometer with a vertically dispersive dipole magnet and a phosphorescent LANEX screen was used to measure energy changes of the probe bunch. Two quadrupole magnets were adjusted such that deflections by transverse fields induced in the channel were canceled in the vertical plane for increased energy resolution, but not completely in the horizontal plane to allow angular deflection measurements of the probe bunch. The offset of the channel, which was varied by a random laser pointing jitter, was measured downstream by imaging the laser profile at multiple object planes using cameras at different distances from the same lens.

%--- THEORETICAL MODEL ---%
The expected wakefields can be modeled by assuming the plasma behaves like a non-evolving dielectric medium \cite{SchroederPRL1999} and that the timescale of the evolution of the beam is long compared to that of the wakefields (quasi-static approximation). Reference \cite{GessnerThesis2016} shows that this results in a single-particle longitudinal wakefield dominated by the fundamental $m=0$ mode, where $m$ denotes the azimuthal index, which is cosine-like in the co-moving longitudinal coordinate $z$,
\begin{equation}
	\label{eq:longwakemodel}
	W_{z0}(z) = - \frac{e k_p \chi_{\parallel}^2 }{2 \pi \epsilon_0 a} \frac{B_{00}(a,b)}{B_{10}(a,b)} \cos(\chi_{\parallel} k_p z) \Theta(z).
\end{equation}
Here $e$ is the positron charge, $\epsilon_0$ is the vacuum permittivity, $k_p$ is the plasma wavenumber, $a$ and $b$ are the channel inner and outer radii, $\Theta(z)$ is the Heaviside step function and
\begin{equation}
	\label{eq:longmodfactor}
	\chi_{\parallel} = \sqrt{\frac{2 B_{10}(a,b)}{2 B_{10}(a,b) - k_p a B_{00}(a,b)}}
\end{equation}
is a longitudinal wavelength modification factor using the ``Bessel-boundary function''
\begin{equation*}
	B_{ij}(a,b) = I_i(k_p a) K_j(k_p b) + (-1)^{i-j+1} I_j(k_p b) K_i(k_p a).
\end{equation*}
The most significant mode of the single-particle transverse wakefield is the sine-like $m=1$ dipole mode
\begin{equation}
	\label{eq:transwakemodel}
	W_{x1}(z) = -\frac{e \Delta x \chi_{\perp}}{\pi \epsilon_0 a^3} \frac{B_{11}(a,b)}{B_{21}(a,b)} \sin(\chi_{\perp} k_p z) \Theta(z),
\end{equation}
whose amplitude is in the direction of the transverse offset $\Delta x$ of the driving particle and where
\begin{equation}
	\label{eq:transmodfactor}
	\chi_{\perp} = \sqrt{\frac{2 B_{21}(a,b)}{4 B_{21}(a,b) - k_p a B_{11}(a,b)}}
\end{equation}
is a transverse wavelength modification factor. Wakefields from arbitrary longitudinal bunch profiles can be obtained by convolving the single-particle wakefield with the particle distribution.

\begin{figure}[t]
	\centering\includegraphics[width=0.95\linewidth]{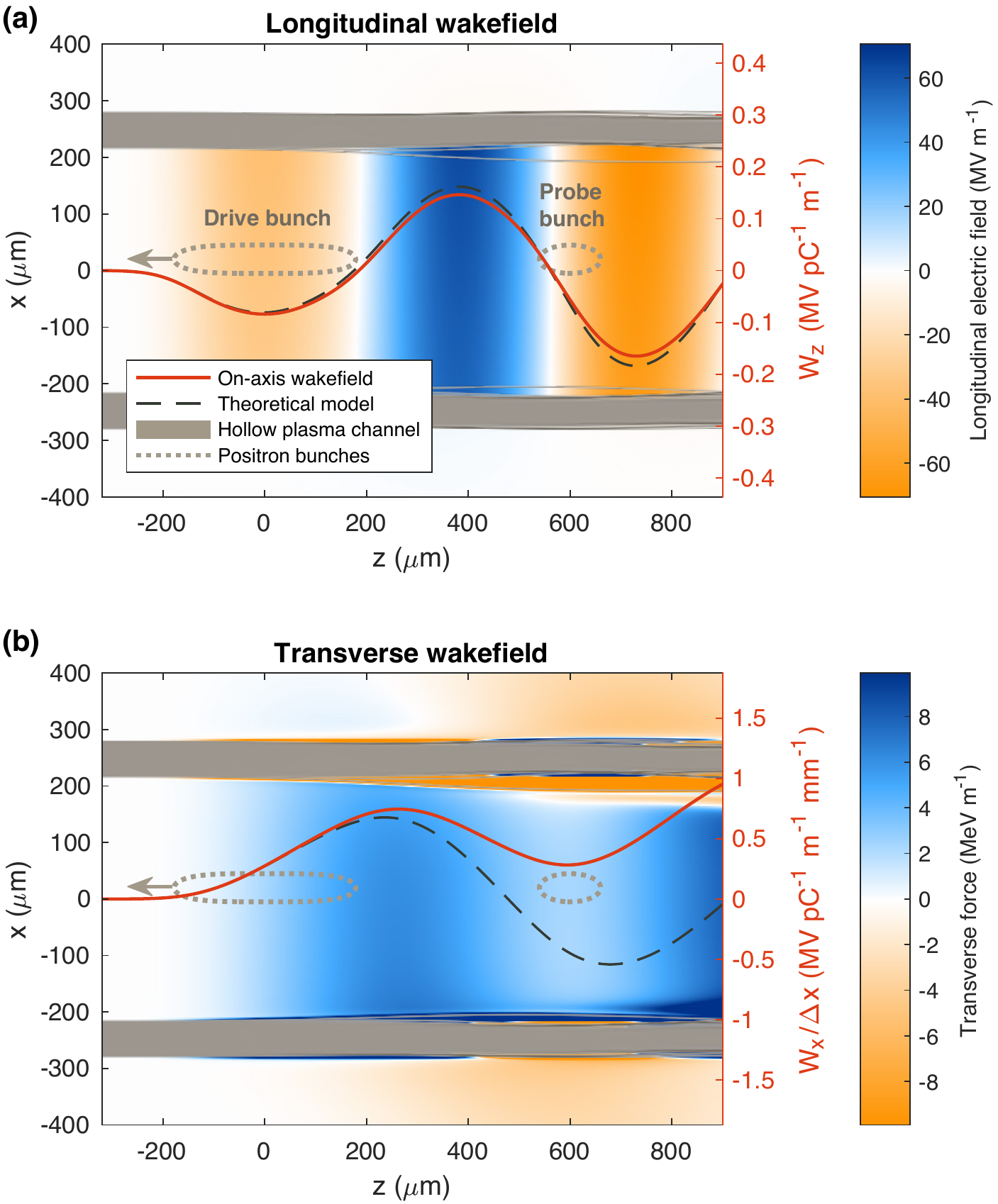}
	\caption{PIC simulation using experimental parameters: a hollow channel with 215~$\mu$m inner and 280~$\mu$m outer radius at density $3\times10^{15}$~cm$^{-3}$, driven by a 0.41~nC drive bunch transversely offset by 20~$\mu$m and probed by a 0.1~nC probe bunch at a bunch separation of 600~$\mu$m. The high beam energy ensures that both the beam and the longitudinal (a) and transverse wakefields (b) stay approximately constant throughout the channel. The on-axis wakefields (red lines) are consistent with the model (black dashed lines) in the longitudinal, but diverges from the modeled transverse wakefield when electrons are pulled into the channel.}
    \label{fig:ChannelQuickPIC}
\end{figure}

%--- SIMULATIONS in QuickPIC ---%
More detailed estimates of the expected wakefields can be obtained from particle-in-cell (PIC) simulations. Figure \ref{fig:ChannelQuickPIC} shows a QuickPIC \cite{AnJCP2013} simulation of a transversely offset beam in a hollow plasma channel using parameters from the experiment. Note the discrepancy between theory and simulation in the transverse wakefield. This is caused by electrons in the wall being pulled into the channel (numerically validated with OSIRIS \cite{FonsecaSpringer2002}), which breaks the assumption of a non-evolving medium.

%--- FIGURE 2: Simulated fields ---%

%--- TRANVSERSE WAKEFIELD from the LONGITUDINAL WAKEFIELD ---%
In addition to a direct measurement, a second independent measurement of the transverse wakefield can be made using the longitudinal wakefield via the Panofsky-Wenzel theorem \cite{PanofskyWenzelRCI1956}, which states that
\begin{equation}
	\label{eq:PWtheorem}
	\frac{\partial W_x}{\partial z} = \frac{\partial W_z}{\partial x}.
\end{equation}
Since the $m=0$ mode of the longitudinal wakefield [Eq.~(\ref{eq:longwakemodel})] cancels due to no $x$-dependence, we must include the much smaller amplitude $m=1$ mode \cite{GessnerThesis2016}
\begin{equation}
	\label{eq:longwakenextmode}
	W_{z1}(z,x) = -\frac{x e \Delta x \chi_{\perp}^2 k_p}{\pi \epsilon_0 a^3} \frac{B_{11}(a,b)}{B_{21}(a,b)} \cos(\chi_{\perp} k_p z) \Theta(z).
\end{equation}
Integrating Eq.~(\ref{eq:PWtheorem}) with respect to $z$ gives to lowest order
\begin{equation}
	W_x(z) = \int_0^z \frac{\partial W_{z1}(z',x)}{\partial x} dz'.
\end{equation}

Since for our parameters $\chi_{\perp} \approx \chi_{\parallel}$, we can relate the $x$-derivative of $W_{z1}$ to the measured $W_z \approx W_{z0}$ by comparing only their amplitudes. This gives the approximate relation
\begin{equation}
	\label{eq:wakeratio}
	\frac{\partial W_{z1}}{\partial x} \approx -\frac{\Delta x}{a^2} \kappa(a,b) W_z,
\end{equation}
where we have simplified the numerical coefficients to
\begin{equation}
	\kappa(a,b) = \frac{4\chi_{\perp}^2 - 2}{\chi_{\parallel}^2 - 1}.
\end{equation}
Finally, we arrive at an equation which allows us to use the longitudinal wakefield to estimate the transverse wakefield per offset,
\begin{equation}
	\label{eq:PWestimate}
	\frac{W_x(z)}{\Delta x} \approx - \frac{\kappa(a,b)}{a^2} \int_0^z W_z(z') dz'.
\end{equation}

%--- INTERPRETATION of DATA ---%
Experimentally, the longitudinal wakefield per particle at the location of the probe bunch $z_{\mathrm{PB}}$ can be determined by the probe bunch energy change $\delta E_{\mathrm{PB}}$, normalized by the charge of the drive bunch $Q_{\mathrm{DB}}$,
\begin{equation}
	W_z(z_{\mathrm{PB}}) = \frac{\delta E_{\mathrm{PB}}}{L_c Q_{\mathrm{DB}}},
\end{equation}
where we have assumed that the channel is uniform along its length $L_c$ and beam loading \cite{KatsouleasPRA1986} is ignored.

\begin{figure}[t]
	\centering\includegraphics[width=0.95\linewidth]{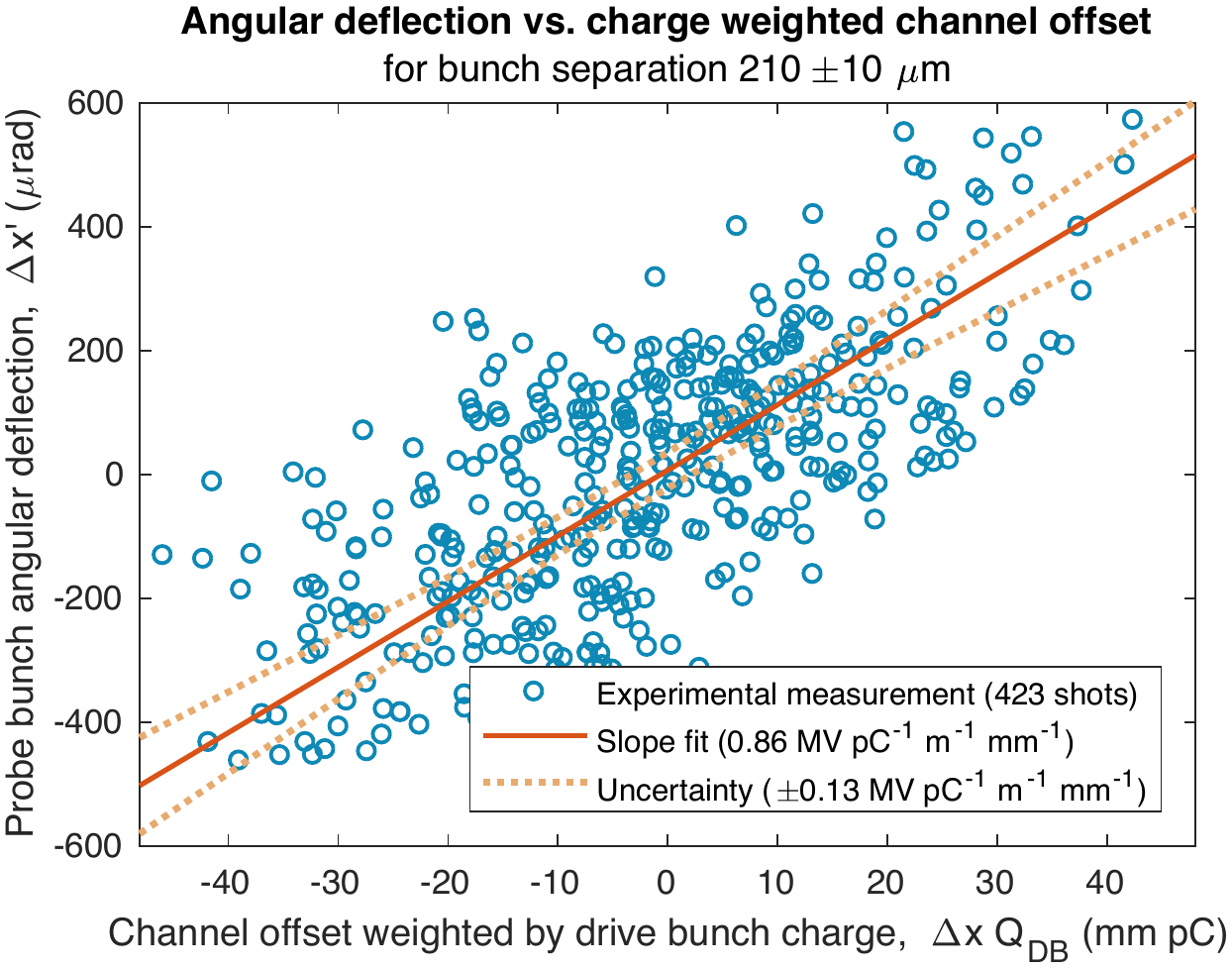}
	\caption{Correlation between probe bunch angular deflection and channel offset weighted by drive bunch charge from a random laser pointing and charge jitter, for the third step ($210\pm10$~$\mu$m) of the bunch separation scan. The linear trend line corresponds to a transverse wakefield $W_x/\Delta x =$~0.86~$\pm$~0.13~MV~pC$^{-1}$~m$^{-1}$~mm$^{-1}$, where the uncertainty is defined by the rms from the trend line increasing by 3\%. The error of each shot is negligible compared to the spread of the data points, caused by a combination of jitters in beam orbit, beam energy, bunch separation, plasma density and channel length.}
    \label{fig:SingleStepKickOffsetCorr}
\end{figure}

Transverse wakefields depend on the transverse offset of the drive bunch. An offset from the channel axis by distance $\Delta x$ drives a transverse wakefield $W_x \propto \Delta x$ [see Eq.~(\ref{eq:transwakemodel})], giving the probe bunch an angular deflection $\Delta x'$. Applying Newton's second law to particles of energy $E_{\mathrm{PB}}$ (large compared to their energy change), we can express the transverse wakefield per particle per offset as
\begin{equation}
	\label{eq:TransWakeEstimation}
	\frac{W_x(z_{\mathrm{PB}})}{\Delta x} = \frac{\Delta x'}{\Delta x Q_{\mathrm{DB}}} \frac{E_{\mathrm{PB}}}{L_c}.
\end{equation}
The slope of the correlation $\Delta x'$ vs $\Delta x Q_{\mathrm{DB}}$ for a large number of shots was measured (see Fig.~\ref{fig:SingleStepKickOffsetCorr}). Note that the offset $\Delta x$ is weighted by the drive bunch charge $Q_{\mathrm{DB}}$ as it varied noticeably across the thousands of shots collected.

The relative beam-channel offset was mainly caused by a random transverse laser jitter of 30-40~$\mu$m rms, measured by laser cameras downstream, whereas the beam orbit in the channel was stable to 5~$\mu$m rms or less. The charge of the drive bunch was determined using the spectrometer upstream of the channel, and the angular deflection of the probe bunch in the horizontal plane as well as its energy change was measured on the spectrometer downstream. For large deflections where the offset was larger than the size of the drive bunch, the probe bunch was also visible on the YAG screen, as seen in Fig.~\ref{fig:ExptSetup}(b). This was used to verify the calibration of the spectrometer angular deflection measurement.

%--- FIGURE 3: Measured transverse wakefield vs bunch separation ---%

\begin{figure}[t]
	\centering\includegraphics[width=0.95\linewidth]{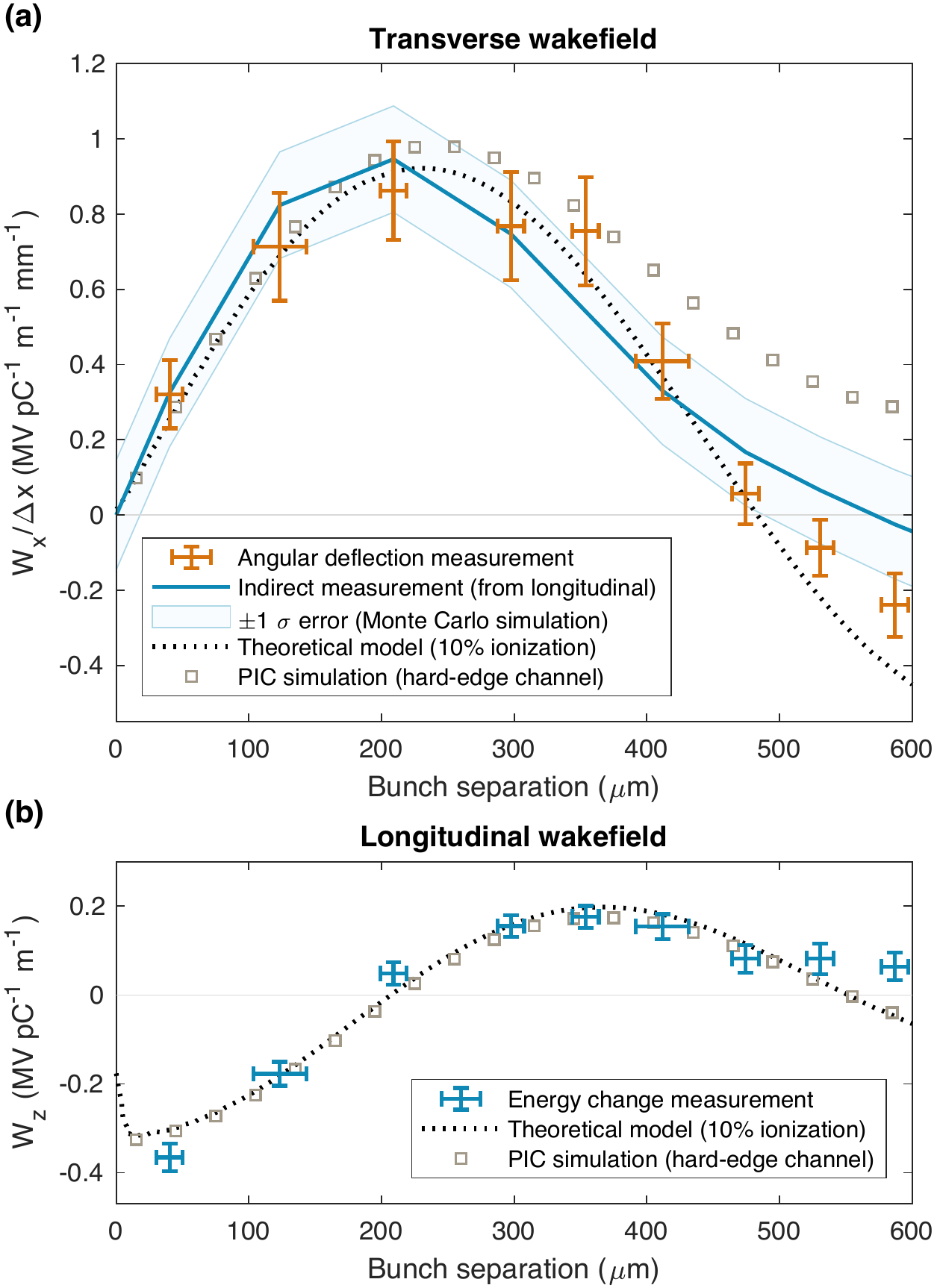}
	\caption{(a) Transverse wakefield from direct measurements (red crosses) and indirectly estimated via the Panofsky-Wenzel theorem (blue line) against bunch separation measured using EOS. Both measurements are initially consistent with theory (dotted black line), but diverge somewhat for larger separations, although not quite matching QuickPIC simulations (gray squares). Notice that the slope in Fig.~\ref{fig:SingleStepKickOffsetCorr} is represented by the third data point. (b) The longitudinal wakefield (blue crosses), largely consistent with theory, is the basis of the indirect transverse wakefield estimate using Eq.~(\ref{eq:PWestimate}). The longitudinal wakefield error, dominated by spectrometer resolution ($\pm$1~pixel), is Monte Carlo simulated to find the indirect measurement error [blue area in (a)].}
    \label{fig:TransAndLongWakefield}
\end{figure}

%--- EXPERIMENTAL RESULTS ---%
Figure \ref{fig:TransAndLongWakefield}(a) shows the measured transverse wakefield per particle per offset for a scan of drive-to-probe bunch separations. The transverse wakefield estimated from the longitudinal wakefield [Fig.~\ref{fig:TransAndLongWakefield}(b)] using the Panofsky-Wenzel theorem is also shown in Fig.~\ref{fig:TransAndLongWakefield}(a) and found to be in good agreement with the measured values. Note that to minimize beam loading effects, only shots with less than 20\% probe-to-drive charge ratio were used to calculate the longitudinal wakefield. The expectation from the theoretical model is found by convolving the single-particle wakefields [Eqs.~(\ref{eq:longwakemodel}) and (\ref{eq:transwakemodel})] with the longitudinal charge distribution measured using EOS. The plasma was found to not be fully ionized, and the plasma density was derived from the wavelength of the measured wakefields, which only depends on the plasma density and the well known radius of the channel. This measurement implies 10\% ionization ($3 \times 10^{15}$~cm$^{-3}$), which is also consistent with known laser parameters. 

Both measurements are largely in agreement with the theoretical model, but diverge somewhat at larger bunch separations. This behavior is expected from the non-linear response of a plasma [see Fig.~\ref{fig:ChannelQuickPIC}(b)], however the measured transverse wakefield does not quite match PIC simulations. We have investigated the effect of more complex radial plasma density profiles, including softer channel walls, but no simulation was found to fully account for the observed discrepancy.

%--- FIGURE 4: Measured and estimated transverse wakefields ---%

%--- DISCUSSION ---%
This measurement shows that a hollow plasma channel generally has the expected transverse wakefield when beams are misaligned with respect to the channel axis. Note however that this is mainly an intra-bunch problem, as the deflection of the accelerated bunch can potentially be canceled by placing it at the zero-crossing of the transverse wakefield (i.e.~close to 500~$\mu$m bunch separation in this measurement). Nevertheless, the issue of transverse deflection of off-axis beams remains, which sets stringent limits on misalignment if used for TeV-scale energy gain. To alleviate this problem, suppression mechanisms must be applied. Suggestions include external focusing or using trains of multiple drive bunches \cite{SchroederPRL1999}, where the longitudinal wakefield is resonantly driven, but the transverse wakefield is not. These and other mechanisms should be further explored to determine whether hollow plasma channels are suitable for high gradient acceleration of positrons.

%--- CONCLUSION ---%
In summary, the transverse wakefield induced by a misaligned positron bunch in a hollow plasma channel has been measured for the first time. These measurements are critical for devising mitigation strategies and alignment tolerances when using hollow plasma channels as accelerating structures.

% --- ACKNOWLEDGEMENTS --- %
\begin{acknowledgments}
The FACET E200 plasma wakefield acceleration experiment was built and has been operated with funding from the United States Department of Energy. Work at SLAC was supported by DOE contract DE-AC02-76SF00515 and also through the Research Council of Norway (Grant No.~230450). Work at UCLA was supported by DOE contract DE-SC0010064 and NSF contract PHY-1415386. Simulations were performed on the UCLA Hoffman2 cluster through NSF OCI-1036224. Simulation work at UCLA was supported by DOE contracts DE-SC0008491 and DE-SC0008316, and NSF contracts ACI-1339893 and PHY-0960344. The work of C.B., S.C.~and A.D.~was supported by the European Research Council (M-PAC project, contract number 715807) and by the France-Stanford Center for Interdisciplinary Studies. The work of W.L.~was partially supported by NSFC 11425521, 11535006, 11175102, and the National Basic Research Program of China Grant No.~2013CBA01501.
\end{acknowledgments}

% --- BIBLIOGRAPHY --- %

\end{document}